# Reducing Hydrocephalus Shunt Revision Rates: A Computational Fluid Dynamics Study on Catheter Hole Design

Omar Said, BSE
Department of Mechanical Engineering
Johns Hopkins University
Baltimore, MD, USA

Mingzi Li, BSE
Department of Biomedical Engineering
Columbia University
New York, NY, USA

Hanyu Gan, BSE
Department of Biomedical Engineering
University of Michigan
Ann Arbor, MI, USA

***Abstract***

Proximal shunt obstruction is the leading cause of ventriculoperitoneal (VP) shunt failure in pediatric hydrocephalus and is closely tied to the near-wall shear environment at drainage holes. We used computational fluid dynamics (COMSOL) to decouple local wall-shear–stress (WSS) control via catheter-tip geometry from global drainage control via a valve opening. A cylindrical ventricle domain with Newtonian CSF ($\rho$ = 1000 kg/m³, $\mu$ = 1 mPa·s) and $\Delta P \approx$ 10 mmHg was solved under laminar Navier–Stokes flow while sweeping distal hole spacing, hole count/diameter, and a simplified valve constriction. Mesh-independence was achieved (minimum element size ≈ 0.021 mm; distal-hole velocity converged to ≈ $1.18\times10^{-3}$ m/s). Perpendicular separation of the two most distal holes increased total outflow by ≈ 22% (2 mm vs 0.5 mm) and raised distal lateral-wall WSS by ≈ 19%, whereas longitudinal shifts of upstream holes produced only small changes. A valve opening near 0.166–0.17 mm set system drainage to ≈ 20 mL/h across geometries, indicating capacity is valve-limited while local WSS is geometry-controlled. The selected tip uses two conical holes (0.5 mm ID) in four rows with 1.4 mm spacing, concentrating WSS at the distal lateral walls, the surfaces where obstruction occurs, without exceeding clinical drainage targets. These findings yield design rules: prioritize distal-pair spacing, reduce hole count with smaller conical openings to elevate protective WSS, and regulate outflow with the valve. The framework provides a clear path to benchtop validation, long-term adhesion assays under mapped WSS, and anatomically realistic CFD.

## I. Introduction

Hydrocephalus is an abnormal buildup of cerebrospinal fluid (CSF) in the brain cavities, called ventricles, caused by an imbalance between production and absorption of CSF. Figure 1 compares the ventricle sizes between people with and without hydrocephalus. There could be an obstruction in the ventricles or inflammations of brain tissues from disease or injury preventing the absorption [1]. Overproduction due to infection or tumor can be a cause as well [2]. The excess fluid widens ventricles and puts pressure on the brain's tissues, which causes symptoms such as headache, sensitivity to light, nausea, and seizures [1,3]. More than 1 million Americans are affected by this medical condition [4].

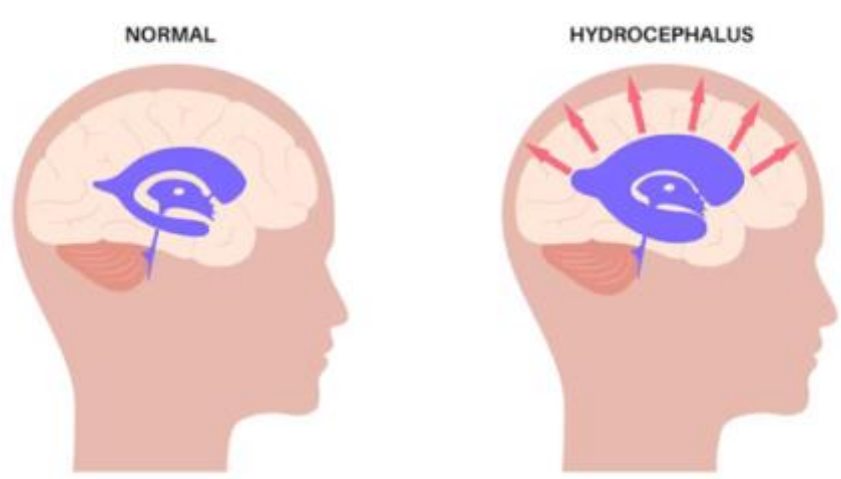

**Fig 1. Hydrocephalus patients have larger ventricles due to excess CSF [4].**

There are largely five types of hydrocephalus: communicating hydrocephalus, non-communicating hydrocephalus, normal pressure hydrocephalus (NPH), acquired hydrocephalus, and hydrocephalus ex-vacuo. Communicating hydrocephalus occurs when the flow of CSF is normal in the ventricles but is blocked after it exits the ventricles. Non-communicating (obstructive) hydrocephalus occurs when the flow of CSF is blocked along narrow passages connecting the ventricles. Normal pressure hydrocephalus (NPH) is a form of communicating hydrocephalus, which occurs more commonly in elderly population. Congenital hydrocephalus is when the patients have the condition from birth. Acquired hydrocephalus is any hydrocephalus condition shown after birth. It could be caused by injuries or diseases. Hydrocephalus ex-vacuo is caused by damage to the brain due to stroke, degenerative diseases like Alzheimer's disease or other dementias or traumatic injury [5].

Currently, there is no definitive treatment without surgery. There are two main treatment methods: endoscopic third ventriculostomy (ETV) and shunt. ETV treats hydrocephalus by making a small hole in the floor of the third ventricle which creates a pathway for CSF to flow normally. ETV is typically used for patients over the age of two with non-communicating hydrocephalus [5].

Placing a shunt is the most common treatment. In the United States, about 125,000 people are living with shunts, and over 36,000 shunts surgeries are performed annually [4, 6]. A shunt is a system that two flexible tubes, also called catheters, with a valve in between run from the ventricles to another body part where CSF can be absorbed naturally such as the abdomen cavity (ventriculoperitoneal; VP), right atrium of the heart (ventriculoatrial; VA), and pleural cavity in lung (ventriculopleural; VPL). VP is the most common type of shunt. A valve connected to the catheter regulates the fluid flow [7]. Since hydrocephalus is a condition caused by excess accumulation of cerebrospinal fluid (CSF) in the ventricles, patients have increased intracranial pressure (ICP) [8]. VP shunt system reduces ICP by draining excess CSF in the ventricles to the abdominal cavity. Figure 2 shows an example placement of a shunt. VP shunt failure rate is about 11-25 % within the first year after initial placement [9]. Pediatric patients (newborn until 21 years old) required more revision surgeries compared to adult patients with rates ranging from 20% to 84.5% in 10 to 15 years of follow-up, and hence, we want to focus on pediatric patients. Pediatric patients required revision surgeries due to their growth, causing misalignment of the shunt. However, shunt obstruction (blockage or occlusion) is the most common malfunction, and the second is infection [9, 10, 11].

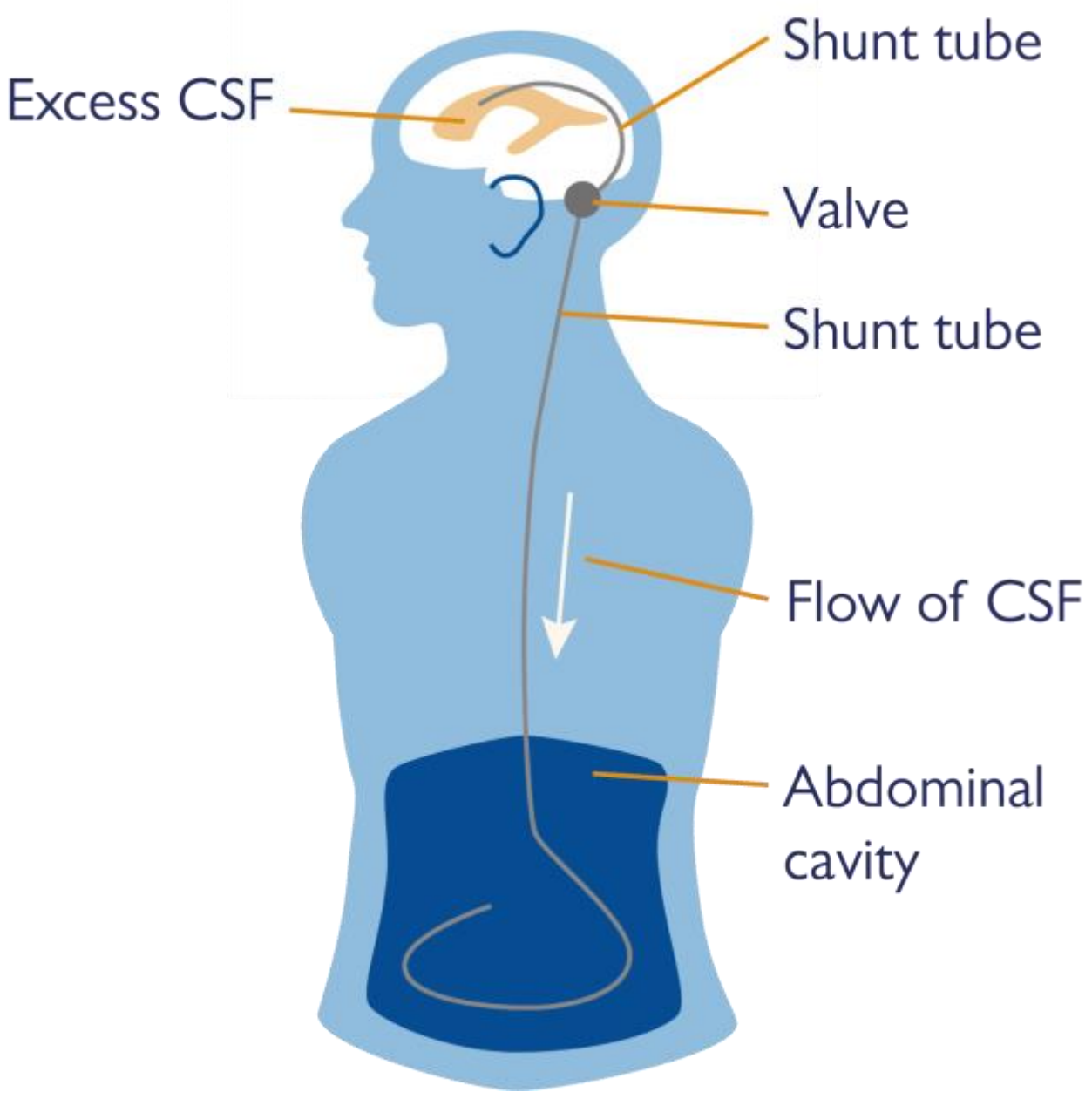

**Fig 2. Shunt tube starts from ventricles and most commonly extends down to abdominal cavity [8].**

Shunt obstruction can occur in the proximal catheter, within the valve, or within the distal catheter. For the proximal catheter occlusion, choroid plexus, a structure producing and maintaining CSF, or migration of brain cells such as astrocytes and microglia, can block the catheter (refer to Figure 3) [12]. Tearing in the brain tissue can cause bleeding, and blood clots can build up in the catheter (refer to Figure 4). Over drainage can pull the ependymal band (ventricular wall) and cause a proximal occlusion [13]. While proximal catheter obstruction is the most common obstruction, there is a study suggesting usage of programmable valves reduces the obstruction [9]. The valve can be blocked due to mechanical failure or clogging with debris or blood clots [12]. Distal catheter occlusion can occur due to adhesions of omentum (a fatty tissue fold in the abdomen) or peritoneal tissues [14].

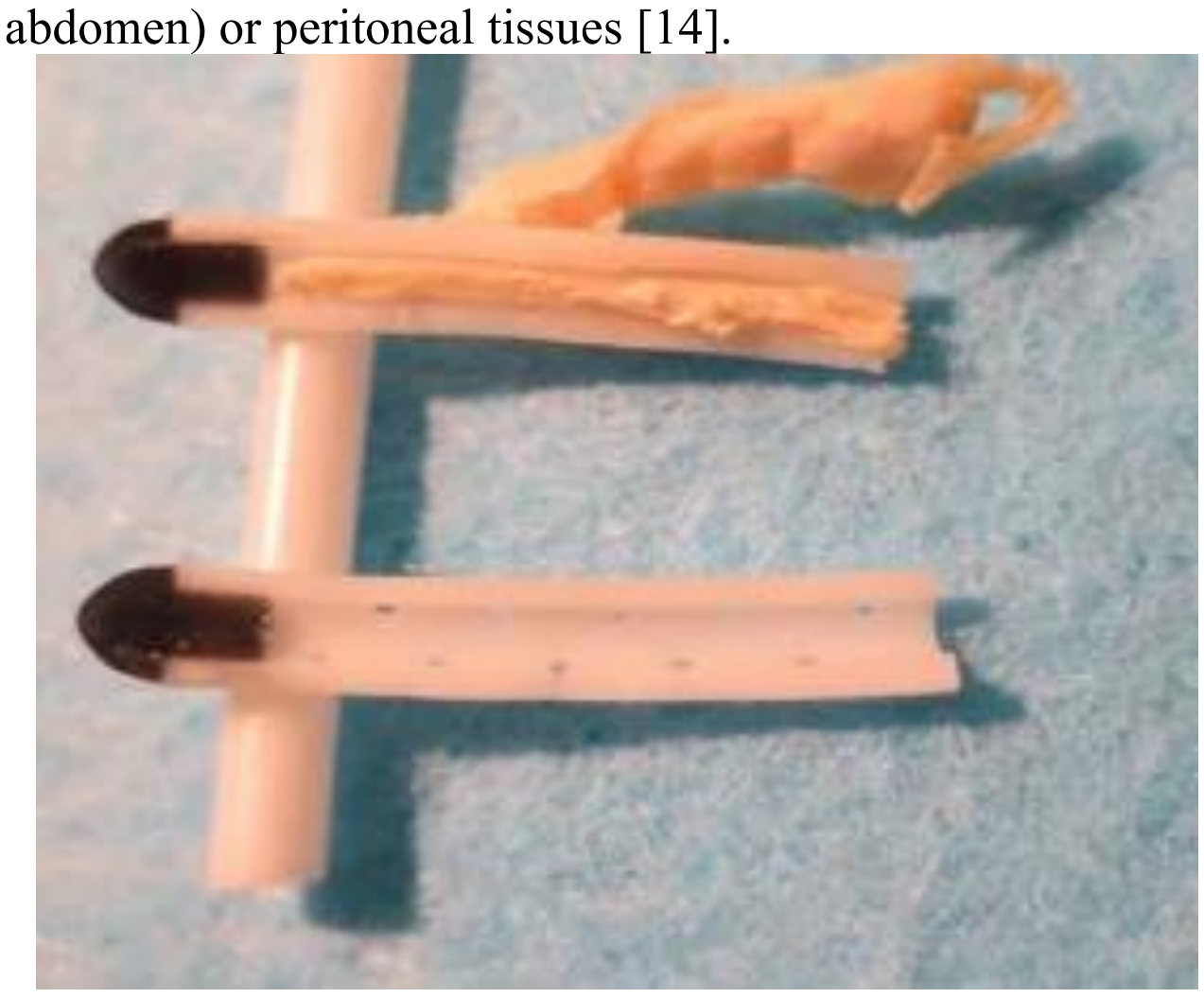

**Fig 3. Brain tissue can completely block the proximal catheter. [15]**

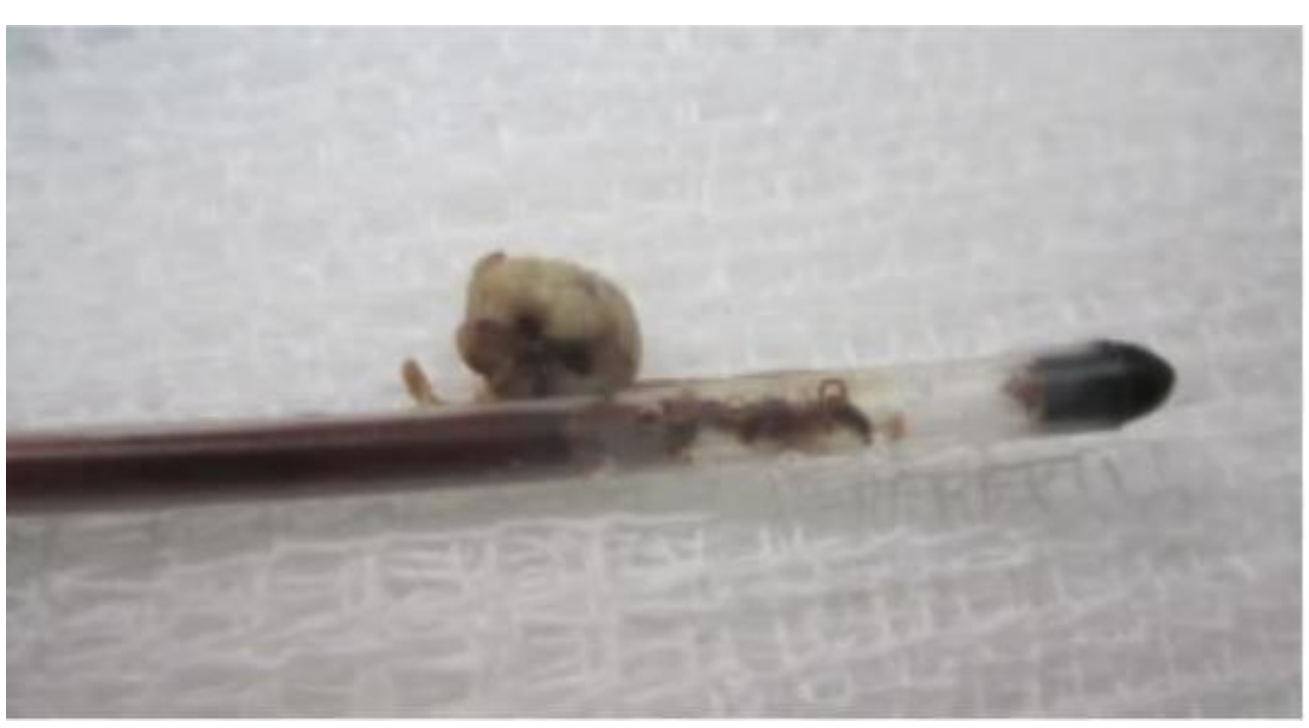

**Fig 4. Blood clots can block the proximal catheter. [15]**

Focusing on distal shunt failure, obstruction often localizes at the catheter slits where CSF exits and arises from tissue ingrowth, proteinaceous deposits, and biofilm formation [9]. Despite improvements elsewhere, distal obstruction remains a persistent challenge and priority for device design and clinical management [16].

Our primary objective in this study is to develop and evaluate a novel catheter hole design specifically for pediatric hydrocephalus patients using computational fluid dynamics (CFD). CFD has been widely used to study physiological fluid flow and disease-related changes in the human body [17]. These computational models can provide insight into flow patterns and mechanical conditions that may be difficult to evaluate experimentally. We will simulate and analyze fluid flow characteristics and catheter performance using COMSOL to optimize the design parameters. This computational approach will establish a foundation for subsequent validation and further research.

## II. Methods

We will be using equations to help model the fluid dynamics in the proximal catheter to understand the flow conditions under different hole designs. F can be set to 0 as external forces will be negligible compared to the forces in the ventricle in hydrocephalus patients. Output velocity, pressure, volumetric flow, and wall shear stress will be solved with COMSOL.

We model the flow in the geometry using the Navier–Stokes equations (Momentum balance equation: balance of forces and Mass balance equation)

$$\vec{u} \cdot \nabla(\vec{u}) = -\frac{1}{\rho}\nabla p + \vec{F} + \frac{\mu}{\rho}\nabla^2\vec{u}, \quad \nabla \cdot \vec{u} = 0$$

Here, $\vec{u} \cdot \nabla(\vec{u})$ represents the convective inertial effects associated with changes in fluid momentum, $-\frac{1}{\rho}\nabla p$ is the pressure-gradient term that drives flow from regions of higher pressure to lower pressure, $\vec{F}$ denotes external body forces such as gravity, and $\frac{\mu}{\rho}\nabla^2\vec{u}$ accounts for viscous diffusion of momentum due to internal friction within the fluid, which resist the flow. The incompressibility condition, $\nabla \cdot \vec{u} = 0$, enforces conservation of mass by requiring the velocity field to remain divergence-free.

To characterize the cerebrospinal fluid flow through the proximal catheter, several derived fluid mechanics quantities were evaluated from the COMSOL solution. The volumetric flow rate through a hole or outlet surface was computed as $Q = \int_A \vec{u} \cdot \vec{n}\, dA$, where $\vec{u}$ is the local fluid velocity and $\vec{n}$ is the outward unit normal to the selected surface. The pressure drop across the system was defined as $\Delta p = p_{\text{in}} - p_{\text{out}}$, where $p_{\text{in}}$is the inlet pressure and $p_{\text{out}}$ is the outlet pressure. Using these values, the hydraulic resistance of the system was evaluated as $R = \Delta p / Q$, which quantifies the opposition of the catheter–valve system to fluid transport. These quantities were used to compare different catheter geometries and to determine how design changes affected drainage performance.

In addition to global flow behavior, local shear conditions at the catheter holes were assessed because wall shear stress is closely related to cell adhesion and possible obstruction. For a Newtonian fluid, the wall shear stress was defined as $\tau_w = \mu \frac{\partial u_t}{\partial n} |_{\text{wall}}$, where $\mu$ is the dynamic viscosity, $u_t$ is the tangential velocity, and $n$ is the direction normal to the wall. This quantity describes the viscous force per unit area exerted by the fluid on the catheter surface. To further justify the laminar flow assumption, the Reynolds number was considered, $Re = \frac{\rho U D}{\mu}$, where $\rho$ is fluid density, $U$ is a characteristic velocity, $D$ is a characteristic diameter, and $\mu$ is dynamic viscosity. Since the Reynolds number in this system remains low under physiological CSF flow conditions, the flow was modeled as laminar throughout the simulations.

All shunt proximal catheter samples included in the simulation also have the same domain setup. Domain 1 represents the ventricle, modeled as a cylinder, shown in Figure 5.

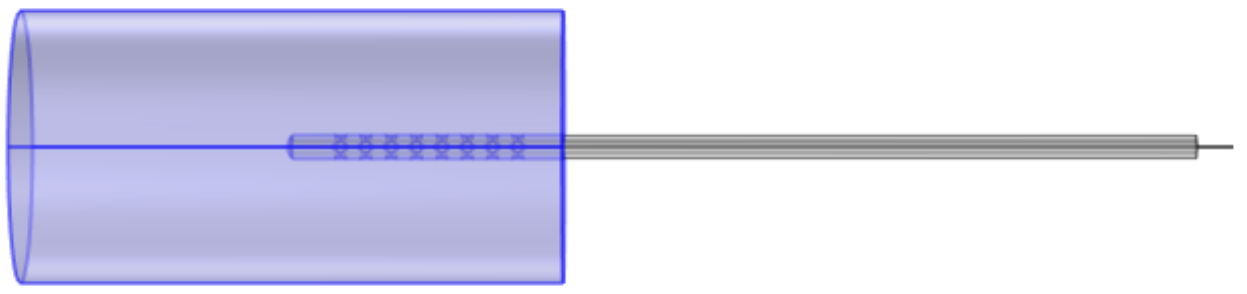

**Fig 5. Domain 1 models the ventricles as a cylinder (height 60mm, basal diameter 30mm)**

Domain 2 represents the proximal catheter, shown in Figure 6.

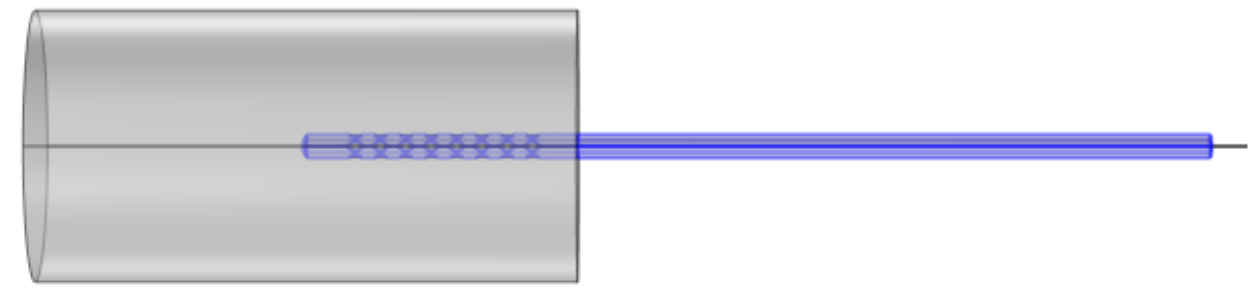

**Fig 6. Domain 2 models the proximal catheter**

Domain 3 represents the lumen of the proximal catheter where the CSF fluid will flow, shown in Figure 7.

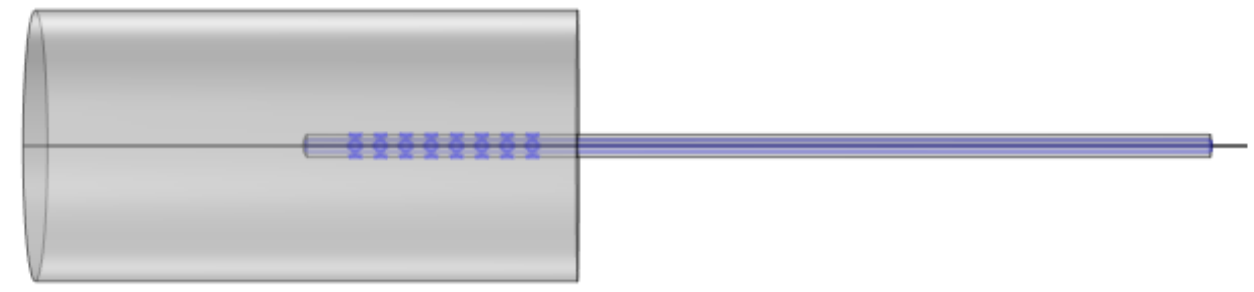

**Fig 7. Domain 3 models the lumen of the proximal catheter where the CSF fluid will flow**

The boundary conditions are how the surroundings and environment affect and interact with the model at its boundaries (edges or surfaces). They specify the conditions at the model's interface with the external environment; in our case, this is the ventricle. The initial conditions define the state of the system at the beginning of the simulation (t = 0). In our model, boundary conditions are used to simulate the laminar flow of CSF fluid through the proximal catheter. Specifically, we set boundary conditions at three locations: the entrance (inlet) at the top of the cylinder (ventricle) and the exit (output) of the proximal catheter end. The catheter walls are set to be non-slipping. The inlet pressure is set to be 10 mmHg to quantify the flow in the ventricle, and the outlet pressure will be set to be 0 as a pressure reference (Table 1in the appendix section).

In the study of the pressure-flow characteristics of the valve, the inlet pressure is subject to change within a range to reflect the pressure increase in a human ventricle resulting from cerebrospinal fluid accumulation. Therefore, the pressure-flow curve could be drawn as the volumetric flow rate changes as a function of the pressure drop.

For our initial design, we adopt the dimensions based on the current ventricular catheter from Medtronic and from the Harris and McAllister [18] study. The conical shape is chosen based on Giménez et al. [19]. The total length is 23 cm, and the length between the tip and the last hole is 1.6 cm. Inner diameter is 1.3 mm, and outer diameter is 2.5 mm. The design is a catheter that has two 0.975 mm diameter holes in four rows, as shown in Figure 8. The choice of 0.975 mm came from a previous study showed minimal growth of astrocytes and macrophages [18]. The total length, length between the tip and the last hole, as well as the inner diameter are aligned with the Medtronic's benchmark product.

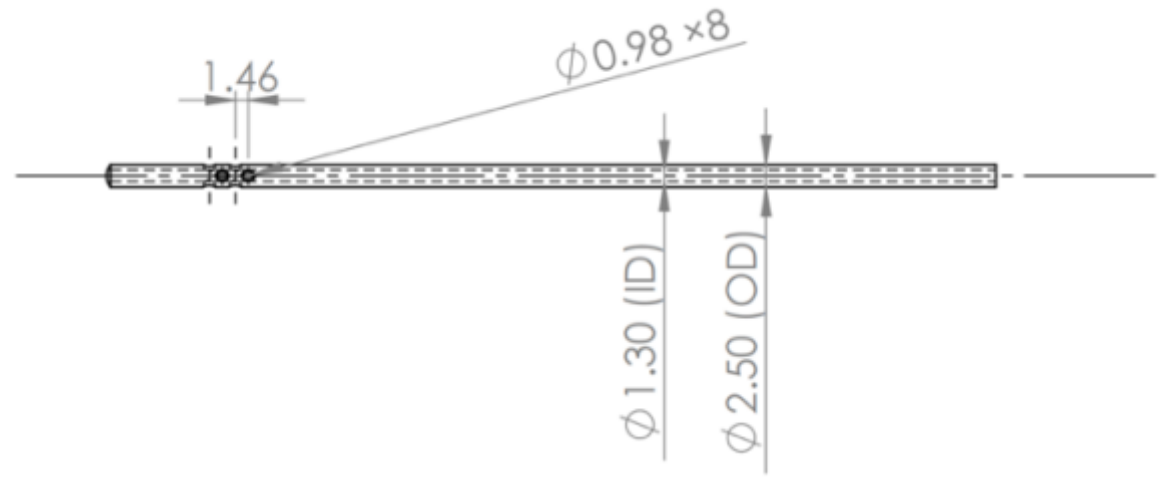

**Fig 8. The consensus design (preliminary ideal prototype) with two conical holes in four rows with a diameter 0.975 mm.**

The lateral surface of one of the two most distal holes to the tip, the counterpart in the most proximal to the tips, and the lateral surfaces of all holes are defined as a selection each. The wall shear stress is measured on all three selections as surface average. Figure 9 shows the location of one of the two most distal holes to the tip, as an example.

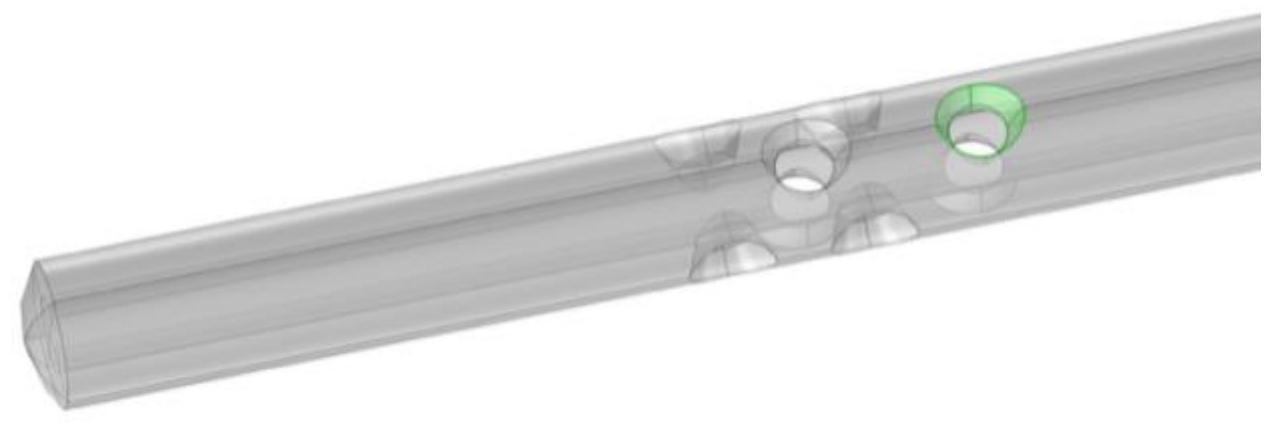

**Fig 9. The location of the lateral walls (on one of the most distal holes to the tip) where wall shear stress is evaluated**

To determine our ideal design, we iterate simulation studies on different designs. To determine if a design was optimal, we applied the fluid mechanics simulations to observe how the fluid would flow through different designs. Our goal is to optimize shear stress and minimize cell growth and tissue ingrowth. This is because cell coverage of drainage holes is considered a function of the wall shear stress. According to Harris and McAllister's procedure, less than or equal to 20% macrophages and less than 10% astrocytes is considered better than the benchmark [18]. If the shear is too high, cell growth is mitigated, but there is too much tissue ingrowth [13]. If the shear is too low, the tissue ingrowth is reduced, but cell growth is not facilitated [18]. Our model focuses on optimizing shear to prevent obstruction to the catheter.

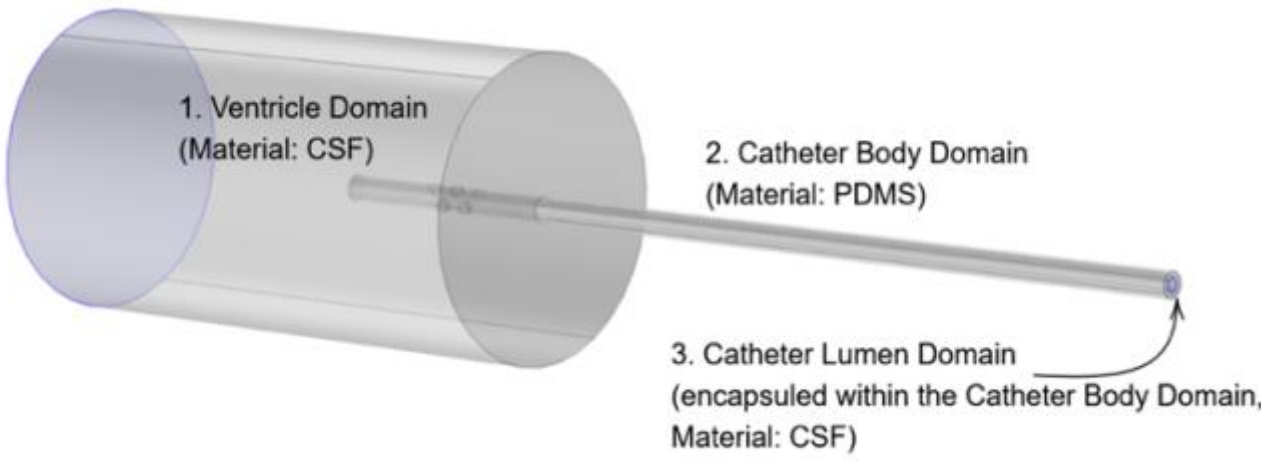

**Fig 10. The domain setting for the ventricle and drainage catheter system**

Three domains are set to validate our catheter design through simulation, which is demonstrated in Figure 10. The domains corresponding to boundary conditions are also included in Figure 11. The inlet is set to show the influence from the choroid plexus, which is the only region generating CSF on the walls of the ventricles. The outlet is set as the distal opening of the ventricular catheter.

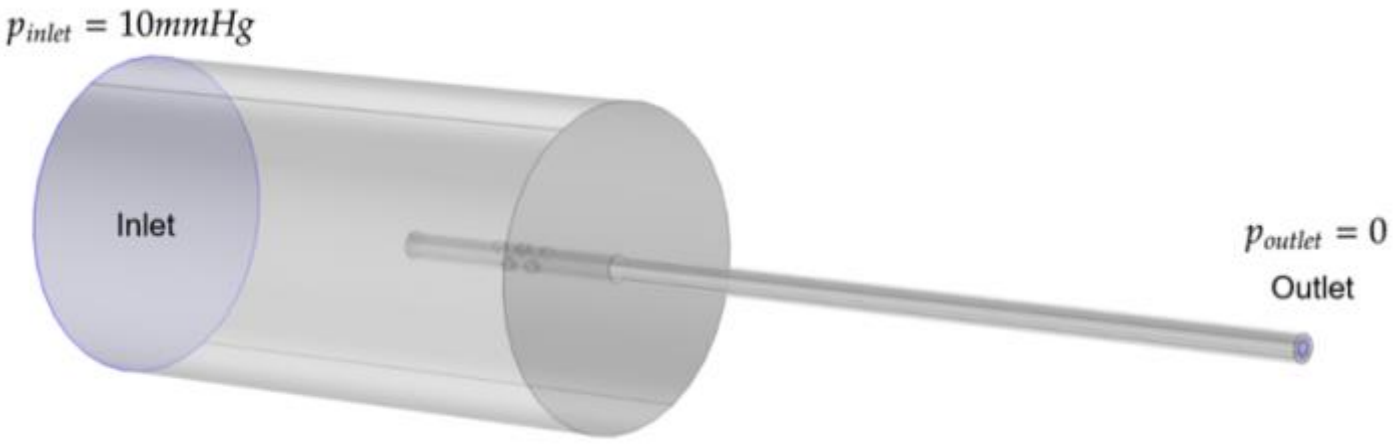

**Fig 11. The boundary conditions for the ventricle and drainage catheter system**

Several parametric sweeps on geometric dimensions are carried out to approach the hole configuration, providing the ideal fluid mechanics property. The specific parameters being swept are summarized in Table 2 in the appendix section.

After more extensive literature research, we locate a more recent paper that combines both astrocyte adhesion experiment under wall shear stress (WSS) factors, and optimization of proximal shunt catheter under fluid simulation and analytical modeling [20]. The results demonstrate different trends in some aspects. Therefore, we decide to make an assumption on the astrocyte adhesion trend under different wall shear stress that higher wall shear stress leads to less cell adhesion [20].

The principle of operation of proximal shunt catheter can be broken down into two independent processes from an engineering perspective; the first is the cell adhesion-wall shear stress relationship, while the second is the wall shear stress-catheter geometry design relationship. The second process is well-understood and explained [20], while the first process is complicated as its dependency on the specific range and threshold, cell types, testing methods and other underlying factors are still a frontier of research. This is to say that even if our assumption is wrong, the fluid mechanics study will still be valid. The only thing that should be changed if so, is the specific threshold on wall shear stress. We believe that this offers our design process a clearer goal, while not posing an issue if progress on the first process involving mechanobiology is made.

We also implemented a shunt valve model in our in-silico design to reflect the strong influence of the valve in the volumetric flow rate of a shunt system in real life. The valve is modeled as a constricted segment of catheter, of which the opening diameter is decided by a parametric sweep on geometry with a goal of reaching a total volumetric flow rate (i.e. drainage rate) of 20 mL/h.

The only difference, between our initial design and this design, is that the introduction of a valve model and the parametric sweep is going to change resistance and Q (seen in Figure 12).

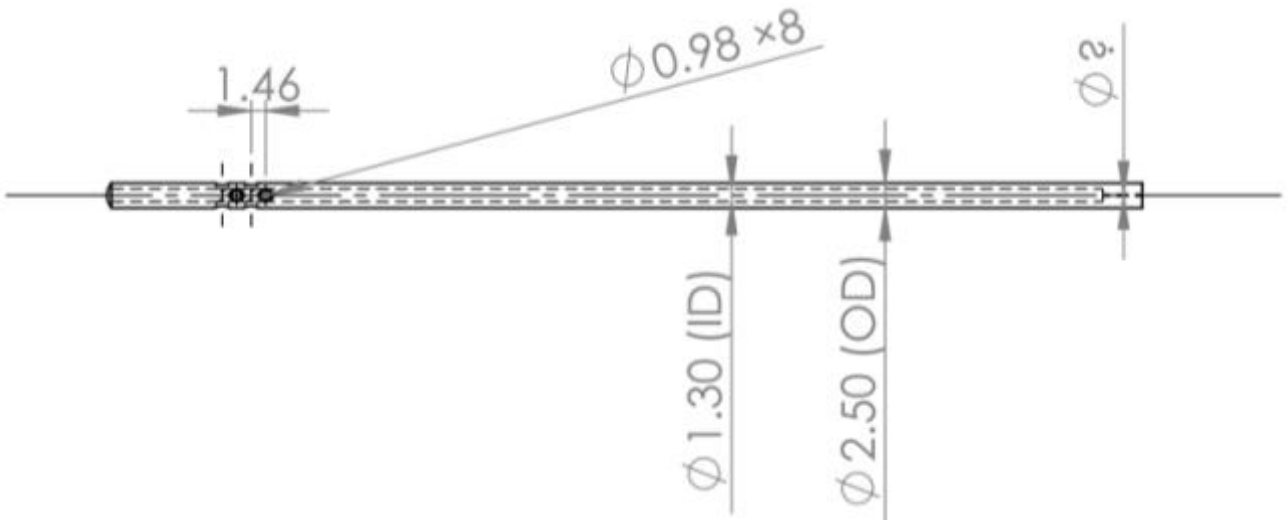

**Fig 12. The same initial design but with a valve (in the downstream of the output). We will determine the final valve diameter via parametric sweep**

A mesh independence study is a critical step in computational simulations to ensure that numerical results are not influenced by the size or density of the computational mesh. In this process, simulations are performed using progressively finer meshes, and key output parameters are monitored for convergence. When the results no longer change significantly with further mesh refinement, the solution is considered mesh independent. This ensures that the simulation accurately represents the physical behavior of the system without being affected by numerical artifacts. Distal hole velocity was selected as the indicator variable for convergence assessment and meshes were refined progressively to evaluate stability of the results.

We will then evaluate the COMSOL CFD simulations by analyzing volumetric flow, wall shear stress, and pressure-flow characteristics. For the volumetric flow, we will verify and validate the size and dimension of the proximal catheter holes. The volumetric flow rate and the flow velocity magnitude are obtained using the "Laminar Flow" module in COMSOL, which simulates the CSF flow from the ventricle to the proximal catheter through the holes. This includes modeling the velocity and the volumetric flow at the distal hole pairs. With these values, we can validate the flow at the hole areas and perform a comparative analysis of the different hole designs.

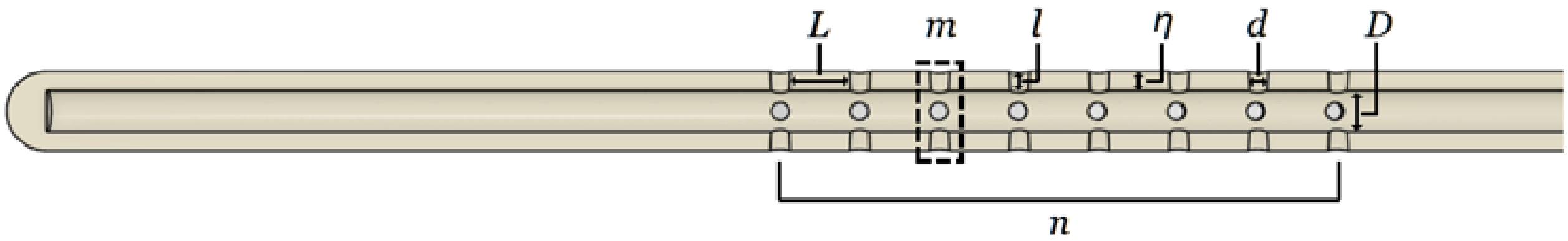

**Fig 13. The dimensions of the catheters explained: Hole diameter d, Lumen diameter D (fixed at 1.3mm), Depth l (fixed at 0.6mm), Wall thickness η (fixed at 0.6mm), Spacing interval L, Each hole segment (dashed box) has m holes (fixed at 4 holes per segment), and the total number of hole segments is n.**

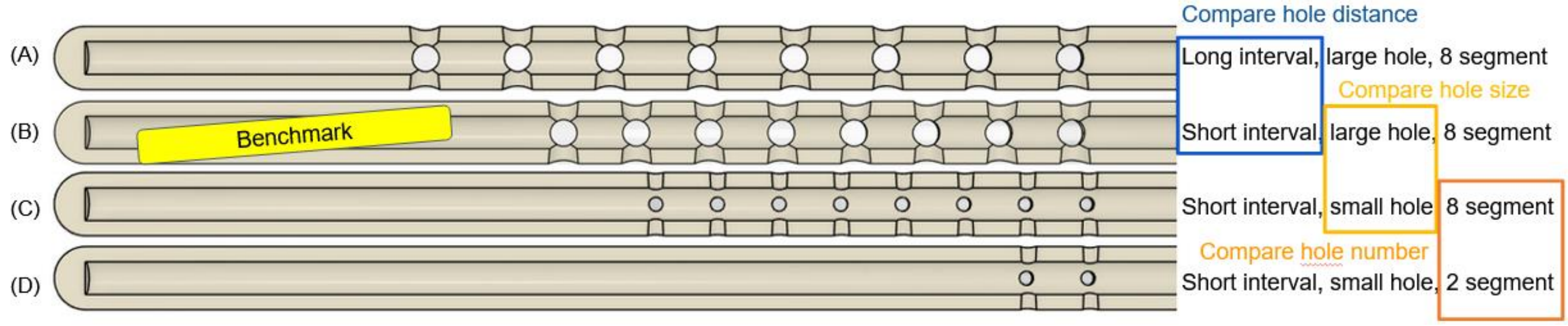

**Fig 14. (A): L = 1.6 mm, d = 0.975 mm, n = 8; (B): L = 0.8 mm, d = 0.975 mm, n = 8; (C): L = 0.8 mm, d = 0.5 mm, n = 8, (D): L = 0.8 mm, d = 0.5 mm, n = 2**

The dimensionless parameters are shown in Figure 13, which are describing the 4 considered designs, with one dimension altering at a time, are included in Table 3 in the appendix section.

There are 4 shunt proximal catheter samples included in the simulation (Figure 14). (B) represents the standard benchmark product manufactured by Integra Codman. This product acts as a baseline positive control in the simulation tests. The other three samples are our proposed adapted prototypes. All 4 shunt proximal catheter samples included in the simulation used the same parameters and input values. The parameters are: inlet pressure of 10 mmHg, CSF density of 1 g/ml, CSF viscosity of 1 mPa*s, system temperature of 310 K, thickness of the catheter wall of 0.6 mm, inner diameter of the catheter of 1.3 mm, longitudinal gap between the center of the adjacent holes of 2.925 mm, and the hole diameter of 0.975 mm for 0.975 mm conical design, and 0.5 mm hole diameter for 0.5 mm conical design. The longitudinal gap between the center of the adjacent holes and the hole diameter parameters are for the parametric sweep on the hole dimensions (sizes and locations).

Convergence was observed around a mesh refinement coefficient of 4, corresponding to a minimum mesh size of 0.021 mm. Beyond this refinement level, further mesh refinements yielded negligible changes in velocity magnitude at the distal hole. This indicates that the chosen mesh resolution provided sufficient accuracy while balancing computational cost. Figure 15 demonstrates the convergence trend, confirming mesh independence of the simulation results.

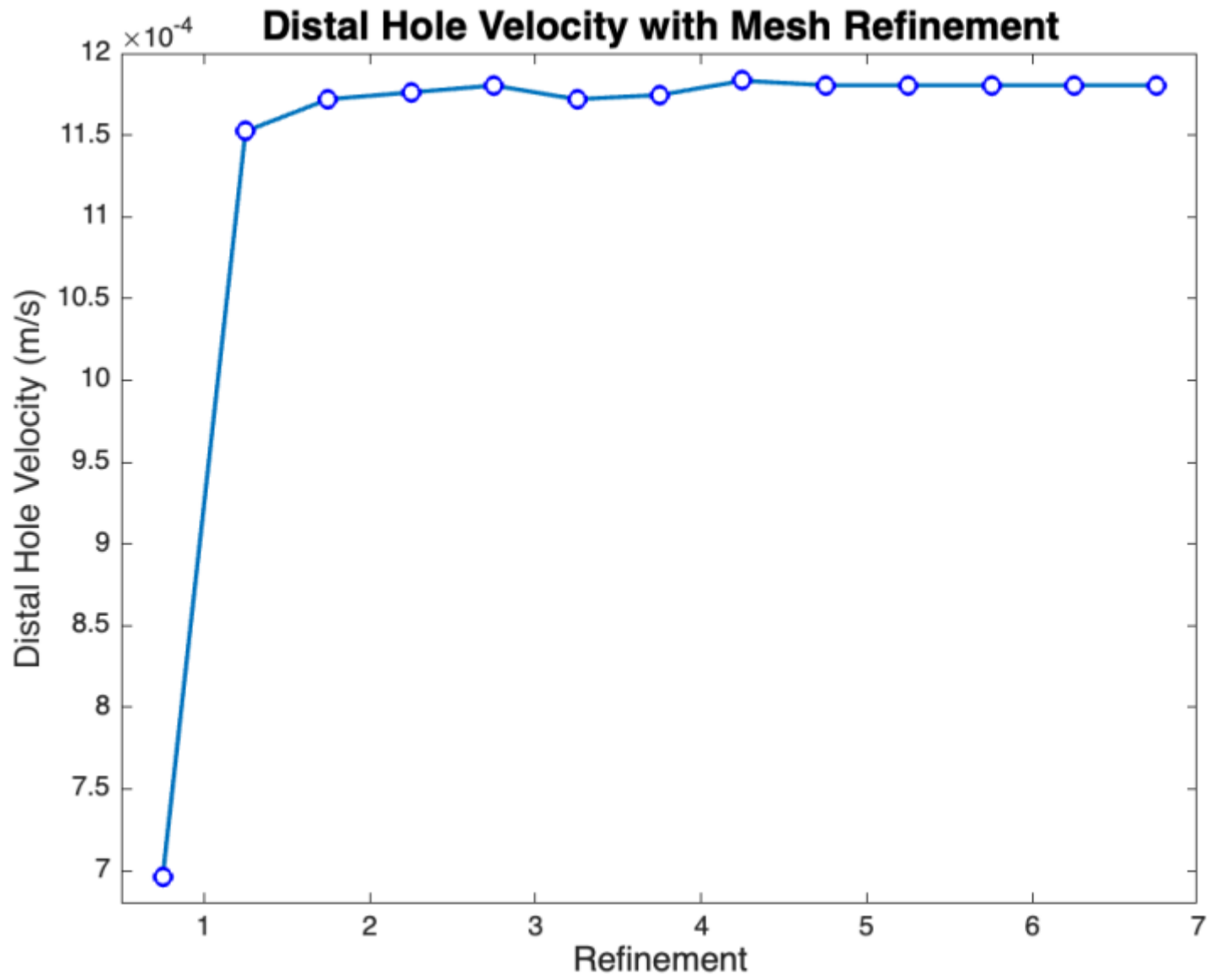

| mesh_refinement | Distal Hole Velocity Magnitude (m/s) |
|---|---|
| 0.75000 | 6.9610E-4 |
| 1.2500 | 0.0011524 |
| 1.7500 | 0.0011720 |
| 2.2500 | 0.0011762 |
| 2.7500 | 0.0011804 |
| 3.2500 | 0.0011720 |
| 3.7500 | 0.0011746 |
| 4.2500 | 0.0011833 |
| 4.7500 | 0.0011806 |
| 5.2500 | 0.0011806 |
| 5.7500 | 0.0011806 |
| 6.2500 | 0.0011806 |

**Fig 15. The graph and table show that as mesh refinement increases, the distal hole velocity converges toward a stable value of approximately $1.1806\times10^{-3}$ m/s, indicating mesh independence beyond a refinement level of 4.**

## III. Results

The results, in summary, shows that changing the hole patterns on the most distal region has the greatest influence on the flow pattern of CSF according to our simulation. As shown by Figure 15, the further the second most distal hole group is away from the most distal hole groups, the stronger the role the most distal holes are going to take on drainage. Quantitatively, the distal holes are going to have larger volumetric flow rates and average velocity magnitudes. At the same time, the wall shear stress on the lateral surface of the distal hole is also increasing, as a result of growing velocity profile.

This offers freedom to adjust the fluid drainage pattern of the distal hole, noted that the maximum drainage rate with a 2 mm perpendicular gap setting ($1.74\times10^{-7}$ $m^3/s$) is 22.2% larger than the setting of 0.5 mm gap ($1.43\times10^{-7}$ $m^3/s$). This is rather significant as the total draining volumetric flow rate is ($5.25\times10^{-7}$ $m^3/s$), pre-valve flow, and this variation accounts for about 6% of the total drainage capacity in our design.

In a similar manner, the average wall shear stress on the distal holes under 2 mm perpendicular gap setting demonstrates a 18.8% increase compared to 0.5 mm perpendicular gap, also showing the capability to tune the property regarding the distal holes. The variation on longitudinal gap shows very little influence on the distal draining behavior (Figure 15), alluding that the proximal holes have limited influence when the distal hole pattern is decided. The videos of the two parametric sweep studies can be seen: https://drive.google.com/file/d/1efNMORo3yxsZ7CMeZeXTgrYD_6_Zga9W/view?usp=sharing (corresponding to Figure 15) and https://drive.google.com/file/d/1qG_-5KUsa7w6879q2EqDe1DR5UKQQKeh/view?usp=sharing (corresponding to Figure 17)

| hole_gap (mm) | Distal Hole Velocity Magnitude (m/s) | Distal Hole Volumetrtic Flow Rate (m^3/s) | Distal Lateral Surface Average for Shear Stress (Pa) |
|---|---|---|---|
| 2.5000 | 0.10734 | 1.6709E-7 | 2.2494 |
| 2.6000 | 0.10845 | 1.6882E-7 | 2.3007 |
| 2.7000 | 0.10856 | 1.6898E-7 | 2.3247 |
| 2.8000 | 0.11015 | 1.7145E-7 | 2.3434 |
| 2.9000 | 0.10927 | 1.7009E-7 | 2.2873 |
| 3.0000 | 0.10801 | 1.6813E-7 | 2.2647 |
| 3.1000 | 0.11023 | 1.7158E-7 | 2.3416 |
| 3.2000 | 0.10920 | 1.6998E-7 | 2.3112 |
| 3.3000 | 0.10892 | 1.6955E-7 | 2.3189 |
| 3.4000 | 0.10932 | 1.7017E-7 | 2.3002 |
| 3.5000 | 0.10803 | 1.6816E-7 | 2.2731 |

**Fig 16. The tabulated results of the parametric sweep on longitudinal gaps of the holes on a proximal catheter**

| perpendicular_gap (mm) | Distal Hole Volumetrtic Flow Rate (m^3/s) | Distal Hole Velocity Magnitude (m/s) | Distal Lateral Surface Average for Shear Stress (Pa) |
|---|---|---|---|
| 0.50000 | 1.4256E-7 | 0.091585 | 1.9813 |
| 0.60000 | 1.5074E-7 | 0.096845 | 2.0694 |
| 0.70000 | 1.5264E-7 | 0.098064 | 2.0954 |
| 0.80000 | 1.5650E-7 | 0.10054 | 2.1738 |
| 0.90000 | 1.5983E-7 | 0.10268 | 2.2035 |
| 1.0000 | 1.6500E-7 | 0.10600 | 2.2263 |
| 1.1000 | 1.6425E-7 | 0.10552 | 2.2167 |
| 1.2000 | 1.6613E-7 | 0.10673 | 2.2721 |
| 1.3000 | 1.6925E-7 | 0.10873 | 2.3324 |
| 1.4000 | 1.6846E-7 | 0.10822 | 2.3065 |
| 1.5000 | 1.7249E-7 | 0.11082 | 2.3837 |
| 1.6000 | 1.7192E-7 | 0.11045 | 2.3311 |
| 1.7000 | 1.7151E-7 | 0.11019 | 2.3611 |
| 1.8000 | 1.7340E-7 | 0.11140 | 2.4007 |
| 1.9000 | 1.7264E-7 | 0.11091 | 2.3836 |
| 2.0000 | 1.7414E-7 | 0.11187 | 2.3529 |

**Fig 17. The tabulated results of the parametric sweep on perpendicular gaps of the holes on a proximal catheter**

Three rounds of parametric sweep on geometry are conducted to finally decide the equivalent valve opening diameter for the shunt system, which is 0.1668 mm under a pressure difference (between choroid plexus and catheter outlet) of 10mmHg. The data are visualized in Figure 18.

Additional gif file can be found in the following links. https://drive.google.com/file/d/17GvlB7U1o2trRoja-iHXynGoj2QtOFZz/view?usp=sharing (pressure) https://drive.google.com/file/d/142yOkc5C3HiqWbngLrqJmfDU1yAKuj7g/view?usp=sharing (velocity)

We also observed the valve parameter sweep across all 4 different geometries. For the flow results, the outlet, which is the end of the valve component, where the CSF fluid is going exit into the distal catheter is selected to be measured.

The velocity magnitude and volumetric flow rate are measured on the outlet as surface average. These values are compared to pass criteria to validate the design requirements (as seen in Figures 19 A-D).

After the dimension of the valve model is fixed in our simulation model, 4 different proximal catheter hole pattern designs are considered in the hole optimization study. The justification of setting the parameters and changing one at a time can be referred to Lee et al [20], as a part of optimization using non-dimensional numbers that characterize the entire system.

Optimizing the dimensionless groups identified geometry E, a new geometry optimized from benchmark D, as the best-performing design, with parameters (m,n,l,L,d,D) = (4,2,0.6mm,0.8mm,0.5mm,1.3mm). This can be shown visually referring to Figure 19. In the figure, the wall shear stress of design labelled by the purple rhombus is among the highest. Meanwhile, the H1 (The most proximal to the tip) also has a relatively high wall shear stress, making both holes have wall shear stress 30 times higher than the wall shear stress in the catheter lumen. In the future, a clearer optimization termination condition should be designed, taking the factors of the possible weakening of the structure's mechanical property when the hole gap is too small or there are too many holes per segment into considerations. A more justified cut-off threshold of the wall shear stress value in the hole that is most proximal to the tip is also expected to be set.

The final design for the prevention of obstruction in the catheter is improved hole geometry. The design is a catheter that has two 0.5 mm diameter holes in four rows, with hole distance (centroid to centroid) of 1.4 mm. The choice of 0.5 mm of two holes per row is based on our simulation result (Figure 20), as it had the highest WSS, supporting least cell adhesion among all designs [20]. The holes are conical in shape with an inner diameter of 0.5 mm. This design will allow for minimal cell adhesion to the drainage holes by changing the shape and size of the holes. This model optimizes wall shear stress distribution during CSF drainage and connects well to the efficacy design input requirement.

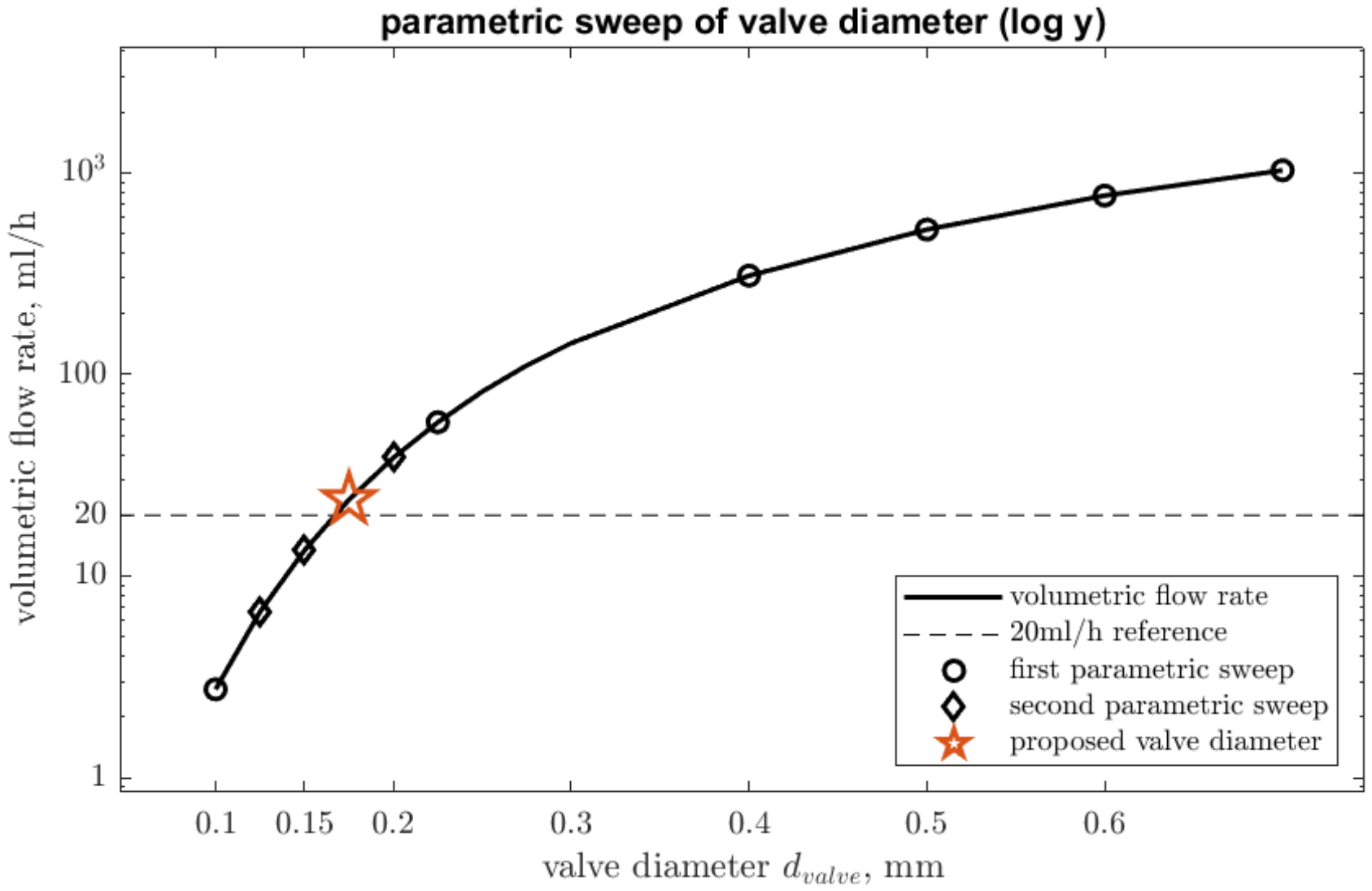

**Fig 18. Effect of valve diameter on the total volumetric flow rate through the shunt system. (y axis is on a log scale)**

| d_valve (mm) | Volumetric Flow Rate (ml/h) | Velocity magnitude (m^3/s) |
|---|---|---|
| 0.16000 | 17.164 | 4.7679E-9 |
| 0.16220 | 18.068 | 5.0188E-9 |
| 0.16440 | 18.964 | 5.2677E-9 |
| 0.16660 | 19.903 | 5.5286E-9 |
| 0.16880 | 20.899 | 5.8052E-9 |

**Fig 19. (A), The first design results**

| d_valve (mm) | Volumetric Flow Rate (ml/h) | Velocity magnitude (m^3/s) |
|---|---|---|
| 0.16000 | 17.164 | 4.7679E-9 |
| 0.16220 | 18.068 | 5.0189E-9 |
| 0.16440 | 18.966 | 5.2684E-9 |
| 0.16660 | 19.904 | 5.5288E-9 |
| 0.16880 | 20.899 | 5.8053E-9 |

**Fig 19. (C), The second design results**

| d_valve (mm) | Volumetric Flow Rate (ml/h) | Velocity magnitude (m^3/s) |
|---|---|---|
| 0.16000 | 17.164 | 4.7679E-9 |
| 0.16220 | 18.068 | 5.0189E-9 |
| 0.16440 | 18.966 | 5.2684E-9 |
| 0.16660 | 19.904 | 5.5288E-9 |
| 0.16880 | 20.899 | 5.8053E-9 |

**Fig 19. (B), The third design, representing the Integra Codman benchmark catheter has the following results**

| d_valve (mm) | Volumetric Flow Rate (ml/h) | Velocity magnitude (m^3/s) |
|---|---|---|
| 0.16000 | 17.183 | 4.7731E-9 |
| 0.16220 | 18.065 | 5.0182E-9 |
| 0.16440 | 18.980 | 5.2723E-9 |
| 0.16660 | 19.947 | 5.5407E-9 |
| 0.16880 | 20.897 | 5.8048E-9 |

**Fig 19. (D), The fourth design, representing our optimized catheter design has the following results**

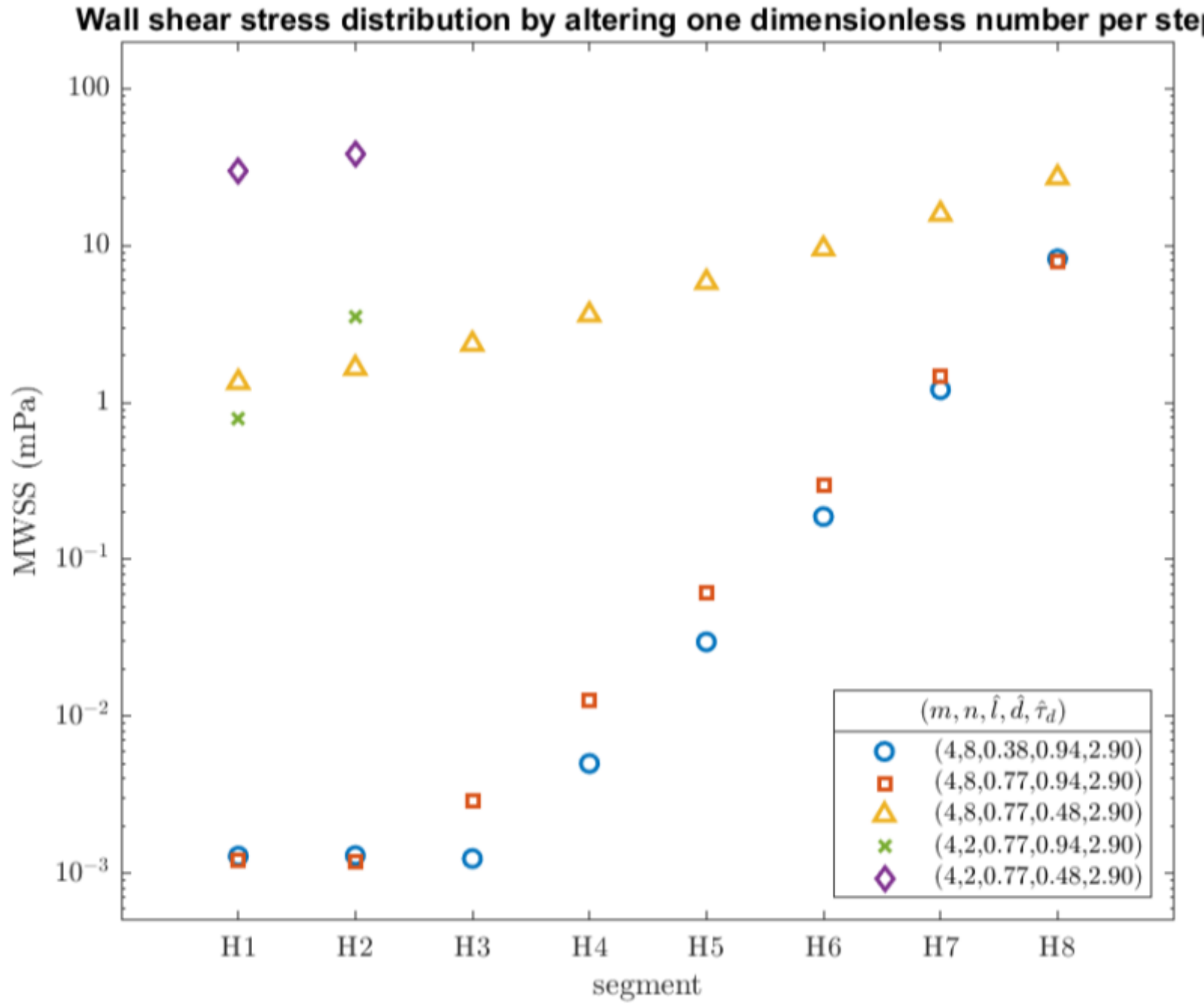

**Fig 20. Wall shear stress distribution by altering one dimensionless number per step. (y axis is on a log scale)**

## IV. Discussion

Before interpreting the results, the final modeled geometries are summarized in Figures 21–23. The catheter domain incorporated conical holes (0.5 mm inner diameter) subtracted from the wall (Figure 21) and a simplified valve orifice appended at the distal end (Figure 22) with an equivalent opening diameter of 0.166 mm ($1.66 \times 10^{-4}$ m). The integrated design (Figure 23) includes two 0.5 mm holes per row across four rows (eight total), with an outer diameter of 2.50 mm, an inner diameter of 1.30 mm, and a 0.17 mm valve segment. The final geometry shown below was used in all simulations and forms the basis for the following discussion on flow distribution, wall shear stress (WSS), and drainage performance.

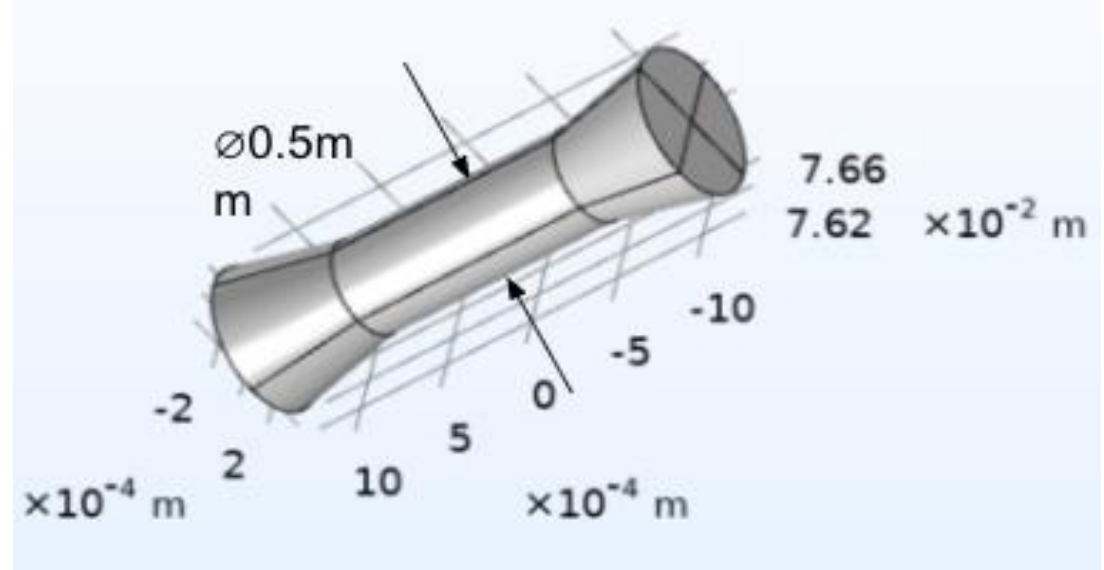

**Fig 21. The hole geometry subtracted from the catheter domain; each conical hole has a 0.5 mm diameter.**

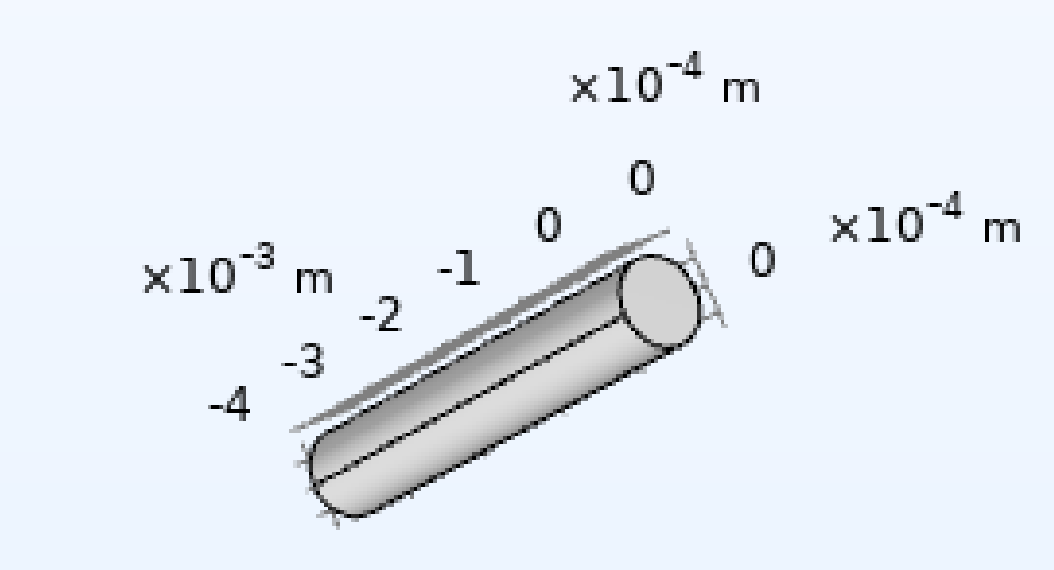

**Fig 22. The valve geometry added to the distal end of the catheter CSF domain; equivalent orifice diameter = 0.166 mm ($1.66 \times 10^{-4}$ m).**

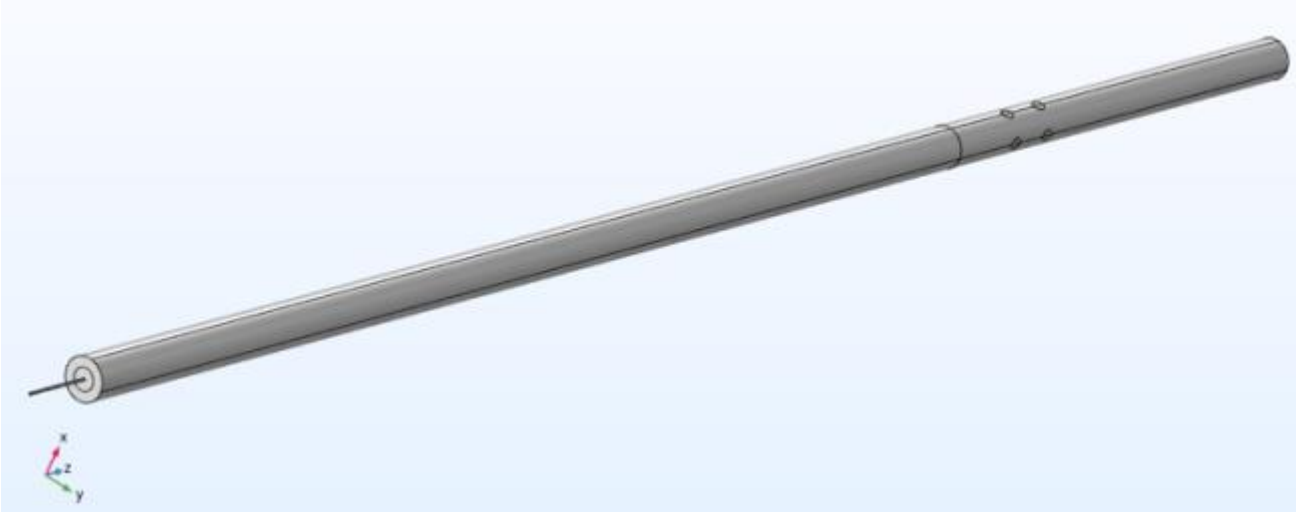

**Fig 23. Final integrated catheter design with two 0.5 mm holes per row, four rows total, OD = 2.50 mm, ID = 1.30 mm, valve = 0.17 mm.**

This work set out to understand the fluid mechanics in the proximal shunt catheter to reduce obstruction risk while preserving clinically appropriate drainage. Four results stand out across studies. Drainage is dominated by the most distal hole group. Moving or spacing these holes alters the volumetric flow and the WSS field more than shifting upstream holes. Longitudinal (tip-to-hole) spacing produced only small changes in capacity (~single-digit percent), whereas perpendicular gap between the paired distal holes produced meaningful differences in both total outflow (≈ +22% at 2 mm vs 0.5 mm) and local WSS (≈+19%) at the distal region.

Geometry can be tuned to raise WSS at surfaces without exceeding drainage targets. Designs that concentrate fewer hole segments and/or employ smaller hole diameters increased WSS on lateral walls near the distal openings, precisely where ingrowth tends to initiate, while the valve provided system-level control to hit a target drainage rate of ~20 mL/h at $\Delta P \approx 10$ mmHg (achieved near a valve opening diameter of ~0.167 mm in our sweep). The numerical conclusions are stable. Mesh refinement demonstrated convergence of distal-hole velocity (≈$1.18\times10^{-3}$ m/s beyond a refinement factor of ~4), supporting that reported differences arise from geometry and boundary conditions rather than numerical artifacts.

Across all four tip geometries, sweeping the valve from 0.1600→0.1688 mm increases total drainage from ≈ 17.16→≈20.90 mL/h, and the corresponding outlet fluxes (≈$4.77\times10^{-9}$→$5.81\times10^{-9}$ $m^3/s$) back-convert exactly to those ml/h values, confirming internal consistency. Crucially, at each fixed valve diameter the spread across designs is negligible (differences typically ≤0.2 – 0.3%: e.g., ~19.90 mL/h at 0.1666 mm for A – C and 19.95 mL/h for D), demonstrating that the valve, not the tip geometry, sets global resistance and clinical drainage in this regime. This supports Result #2: we can decouple objectives, use geometry to raise WSS at the distal lateral walls (where obstruction occurs) without risking under/over-drainage, because the valve reliably tunes total flow to ~20 mL/h at $\Delta P \approx 10$ mmHg. In other words, Fig. 18A – D shows capacity is valve-limited, while our geometry sweeps (distal pair spacing, hole count/size) primarily reshape local WSS, the quantity that matters for adhesion.

Together, these findings suggest a catheter-plus-valve design rule; use geometry to prioritize WSS where occlusion originates (distal holes/lateral walls) and use the valve to regulate the global outflow.

Proximal obstruction remains the leading cause of shunt malfunction, commonly attributed to tissue ingrowth (ependymal bands, choroid plexus) and clot formation. The biophysical lever available in catheter design is the near-wall shear environment. Higher WSS decreases adhesion and early biofilm maturation. Our simulations indicate that distal holes dominate both flow and WSS. This aligns with the clinical observation that flow preferentially leaves nearest the pressure drop and unobstructed openings. It also reframes the design problem, the distal pair does not simply "share" the load; it sets the shear landscape for the entire tip region. Perpendicular separation of the distal pair is a high-leverage knob. Increasing this distance both lifts WSS on each lateral wall and raises total throughput. In contrast, shifting holes along the axis (longitudinal gap) perturbs streamlines minimally once the distal pair is fixed, explaining the small effect on total capacity.

Fewer hole segments with appropriately sized holes channel more of the pressure drop through each opening, raising local velocities and WSS while reducing the number of potential occlusions. This must be balanced against redundancy, too few openings can make the system prone to partial occlusion.

The results motivate a set design principle that prioritize the distal pair. Treat the two most distal openings and their lateral walls as the primary anti-occlusion surfaces. Specify a larger perpendicular gap (≈2 mm) to (i) increase total drainage, (ii) raise WSS on the lateral walls, and (iii) reduce recirculation pockets between the jets.

Also motivates using fewer segments with redundancy. Reducing the number of hole segments concentrates flow (raising WSS) and simplifies manufacturing and inspection. Maintain redundancy via paired holes per row and valve-regulated flow rather than many low-shear openings. Also the design should use conical holes with smaller inner diameters near ~0.5 mm will elevate local shear while the valve ensures that overall system resistance still meets drainage targets.

Additionally, the design should incorporate a valve. Geometry sets the distribution of shear and the valve sets the magnitude of flow. Our sweep shows a practical orifice near 0.166–0.17 mm meets ~20 mL/h at 10 mmHg, leaving geometric degrees of freedom available to maximize distal WSS without risking over drainage. Finally, the design will be validated with surface-average and peak WSS. Because adhesion thresholds are non-linear and may depend on local peaks at hole edges, both average and maximum WSS should be tracked as acceptance criteria.

We adopted the widely used engineering heuristic that higher WSS correlates with lower early adhesion for macrophages and astrocyte processes, acknowledging that exact thresholds vary with cell state, protein conditioning layers, and time under shear. The pattern we observe, distal-biased, higher-shear outlets, is biologically plausible for reducing early ingrowth. One important detail to note, too high a local shear could theoretically irritate adjacent tissue (ependymal surface) or draw in debris if the tip is positioned too close to the wall. Our geometry reduces this risk by using paired distal holes and by shaping holes conically to smooth entrance effects.

In a field where pediatric patients often face multiple revisions within the first years of life, even small increases in time-to-occlusion are meaningful. In this context, catheter geometry alone is insufficient; flow regulation at the system level becomes essential, A programmable or sized valve orifice brings drainage onto a target (~20 mL/h) under typical intracranial pressures, helping prevent both underdrainage and over drainage-driven issues. Also using fewer, higher-shear distal openings helps prevent early blockage from redirecting flow into slow-flow regions where material can build up and spread, a pattern seen in multi-segment designs.

Mesh-independence testing demonstrated stable distal-hole velocities beyond a refinement factor of ~4 (minimum element size ≈ 0.021 mm). Boundary conditions (no-slip walls, $\Delta P \approx 10$ mmHg, Newtonian CSF at 1 mPas, 310 K) reflect standard assumptions for laminar ventricular flow, where Reynolds numbers are low. While the ventricle was idealized as a cylinder and the valve as a simple orifice element, the comparative conclusions (which geometry raises WSS and how spacing affects outflow) rely on relative changes under identical conditions.

There are several limitations in this study: (1) A cylindrical ventricle neglects the complex curvature, choroid plexus morphology, and proximity of ependymal surfaces. (2) We imposed a steady pressure difference; real ICP varies with posture, respiration, and cardiac cycle. Transients can change instantaneous WSS and valve behavior. (3) We did not perform fluid–structure interaction with catheter compliance or wall motion, nor did we include particulate transport or debris capture. (4) WSS is used as a mechanistic indicator of adhesion risk; real outcome depends on protein adsorption kinetics, cell phenotype, biofilm maturation, and time, none of which are explicitly modeled here. (5) Our sweeps varied hole spacing, count, and diameter in structured ways; interactions between factors beyond those tested could yield alternative results (e.g., tapered lumen, staggered oblique holes, micro-textures).

However, this work could be built on by pursing the following steps, Printing catheter tips with the studied geometries, drive synthetic CSF under controlled ΔP and posture, and measure hole-resolved flow (μ-PIV) and WSS surrogates. Then compare the benchwork to this numerical study. Secondly, exposing tips to protein-conditioned flow with astrocyte/macrophage cocultures under mapped WSS levels, then quantify coverage vs WSS and identify practical thresholds. Third, Place the tip inside subject-specific ventricular reconstructions to observe wall interactions, posture effects, and anti-siphon components. Finally, introduce partial occlusion in silico (e.g., single-hole blockage) to test resilience of different patterns and to tune number of holes.

## V. Conclusion

This study links proximal shunt catheter geometry to its fluid-mechanical performance and offers concrete design rules to mitigate obstruction. Across simulations, drainage and WSS patterns were governed by the most distal hole pair, increasing their perpendicular separation raised total flow (~22% at 2 mm vs 0.5 mm) and boosted distal-wall WSS (~19%), while longitudinal shifts of upstream holes had minimal effect. Designs with fewer hole segments and smaller, conical holes (~0.5 mm ID) concentrated flow to elevate protective WSS without exceeding clinical drainage when paired with a valve orifice near 0.166 – 0.17 mm, which achieved ~20 mL/h at $\Delta P \approx 10$ mmHg. Mesh-refinement results confirmed numerical stability, indicating that these trends reflect geometry, not discretization.

By integrating catheter geometry sweeps, valve-level flow control, and mesh-verified CFD, this study clarifies which knobs matter for proximal shunt performance and how to turn them. The most distal holes govern both drainage and the local shear environment where obstruction begins. Increasing their perpendicular separation and reducing hole count/diameter raises protective WSS without sacrificing target outflow when paired with an appropriately sized valve.

Practically, we recommend prioritize the distal pair, maximize their perpendicular gap, reduce hole count with conical ~0.5 mm openings, and regulate system outflow via the valve. Limitations include idealized anatomy, steady Newtonian flow, and absence of biological kinetics. Next steps are benchtop μPIV validation, long-term adhesion assays under mapped WSS, and anatomically realistic CFD. Taken together, the results provide a translational path to simpler, higher-shear, valve-tuned catheters aimed at extending shunt longevity.

## VII. Appendix

| Description | Expression | Value |
|---|---|---|
| The pressure inlet | 135.95 [mmH2O] | 1333.2 Pa |
| The density of CSF fluid | 1 [g/ml] | 1000 $kg/m^3$ |
| The viscosity of CSF fluid | 1 [mPa·s] | 0.001 Pa·s |
| The temperature of system | 310 [K] | 310 K |
| Thickness of the catheter wall | 0.6 [mm] | 6.0E-4 m |
| Inner diameter of the catheter | 1.3 [mm] | 0.0013 m |
| Stiffness of material | 6 [MPa] | 6.0E6 Pa |
| Gap between center of adjacent holes longitudinally | 2.925 [mm] | 0.002925 m |
| Diameter of the holes | 0.975 [mm] | 9.75E-4 m |

**Table 1. The parameters, initial and boundary conditions for fluid mechanics simulation of a ventricle and a drainage catheter placed inside**

| Sweeping Parameter | Sweeping Range | Graphical Illustration | Comments |
|---|---|---|---|
| For each horizontal and vertical hole array, the **longitudinal gap** between the proximal hole and the distal hole is swept (the most distal hole on each array is fixed in position) | Start: 0.5 mm,<br>Step: 0.1 mm,<br>End: 2 mm | | This sweep is to examine the effect of the proximal hole on each longitudinal array. |
| Only the most distal hole on each vertical array is swept regarding position; its **perpendicular gap** between the most distal holes on the horizontal direction is swept | Start: 0.5 mm,<br>Step: 0.1 mm,<br>End: 2 mm | | This sweep is to evaluate the effect of the second most distal holes on the draining of the most distal holes. |

**Table 2. The summary for parametric sweeps of the geometric features**

| Geometry # | $m$ | $n$ | $\hat{l} = \frac{l}{L}$ | $\hat{d} = \frac{d}{D}$ | $\hat{\tau}_d = \frac{32\mu Q}{\pi D^3 \tau_c}$ |
|---|---|---|---|---|---|
| 1 | 4 | 8 | 0.38 | 0.94 | 2.90 |
| 2* | 4 | 8 | **0.77** | 0.94 | 2.90 |
| 3 | 4 | 8 | 0.77 | **0.48** | 2.90 |
| 4** | 4 | 2 | 0.77 | 0.94 | 2.90 |
| 5*** | 4 | **2** | 0.77 | 0.48 | 2.90 |

*: Geometry 2 is the Integra Codman benchmark product
**: Geometry 4 shows the consensus design (preliminary ideal prototype)
***: Geometry 5 shows the most current prototype, optimized from Geometry 3

**Table 3. The hole pattern geometry of the proximal shunt catheter**